\newcommand{\auu}{\mbox{$\overline{\forall}  $}} 
\newcommand{\aee}{\mbox{$\overline{\exists}$}}  
\newcommand{\acc}{\mbox{$\overline{\wedge}$}}  
\newcommand{\add}{\mbox{$\overline{\vee}$}}    
\newcommand{\fff}{\mbox{$\stackrel{}{\bot}$}}          
\newcommand{\ftt}{\mbox{$\stackrel{}{\top}$}}          
\newcommand{\fii}{\mbox{$\stackrel{}{\rightarrow}$}}          
\newcommand{\fnn}{\mbox{$\stackrel{}{\neg}$}}          
\newcommand{\fcc}{\mbox{$\stackrel{}{\wedge}$}}          
\newcommand{\fdd}{\mbox{$\stackrel{}{\vee}$}}          
\newcommand{\feq}{\mbox{$\stackrel{}{\leftrightarrow}$}} 
\newcommand{\fuu}{\mbox{$\stackrel{}{\forall}$}}          
\newcommand{\fee}{\mbox{$\stackrel{}{\exists}$}}          
\newcommand{\pff}{\mbox{$\stackrel{{\bf .}}{\bot}$}}          
\newcommand{\ptt}{\mbox{$\stackrel{{\bf .}}{\top}$}}          
\newcommand{\pii}{\mbox{$\stackrel{{\bf .}}{\rightarrow}$}}          
\newcommand{\pnn}{\mbox{$\stackrel{{\bf .}}{\neg}$}}          
\newcommand{\pcc}{\mbox{$\stackrel{{\bf .}}{\wedge}$}}          
\newcommand{\pdd}{\mbox{$\stackrel{{\bf .}}{\vee}$}}          
\newcommand{\peq}{\mbox{$\stackrel{{\bf .}}{\leftrightarrow}$}} 
\newcommand{\puu}{\mbox{$\stackrel{{\bf .}}{\forall}$}}          
\newcommand{\pee}{\mbox{$\stackrel{{\bf .}}{\exists}$}}          
\newcommand{\rff}{\mbox{$\stackrel{{\bf ..}}{\bot}$}}          
\newcommand{\rtt}{\mbox{$\stackrel{{\bf ..}}{\top}$}}          
\newcommand{\rii}{\mbox{$\stackrel{{\bf ..}}{\rightarrow}$}}          
\newcommand{\rnn}{\mbox{$\stackrel{{\bf ..}}{\neg}$}}          
\newcommand{\rcc}{\mbox{$\stackrel{{\bf ..}}{\wedge}$}}          
\newcommand{\rdd}{\mbox{$\stackrel{{\bf ..}}{\vee}$}}          
\newcommand{\ruu}{\mbox{$\stackrel{{\bf ..}}{\forall}$}}          
\newcommand{\ree}{\mbox{$\stackrel{{\bf ..}}{\exists}$}}          
\makeindex

\documentstyle{article}

\def\noheaderplainsetup{

\topmargin=0pt \headheight=0pt \headsep=0pt  \oddsidemargin=0.5truein \evensidemargin=0.5truein  \textheight=8.9truein \textwidth=5.5truein}   

\noheaderplainsetup

\begin{document}
\newtheorem{theorem}{Theorem}[section]
\newtheorem{claim}[theorem]{Claim}
\newtheorem{convention}[theorem]{Convention}
\newtheorem{notation}[theorem]{Notation}
\newtheorem{definition}[theorem]{Definition}
\newtheorem{lemma}[theorem]{Lemma}
\newtheorem{remark}[theorem]{Remark}
\newtheorem{fact}[theorem]{Fact}
\newtheorem{corollary}[theorem]{Corollary}
\newtheorem{termnot}[theorem]{Terminology and notation}


\title{On resources and tasks}
\author{Giorgi Japaridze\\ 
 \\  \\ Department of Computing Sciences, Villanova University, 
\\ 800 Lancaster Avenue, Villanova, PA 19085, USA
}
\date{}
\maketitle

\begin{abstract} Essentially being an extended abstract of the author's 1998 PhD thesis, this 
paper introduces an extension of the language of linear logic with a semantics
which treats sentences as {\em tasks} rather than true/false statements.
A resource is understood as an agent capable of accomplishing the task
expressed by such a sentence. It is argued that the corresponding logic can be used as a
planning logic, whose advantage over the traditional comprehensive
planning logics is that it avoids the representational \index{frame
problem}frame problem and significantly alleviates the inferential frame problem.
\end{abstract}

\

\

Dedicated  to the dear memory of {\bf Leri Mchedlishvili} (1936-2013):  my first teacher of logic, colleague, and a prominent figure in Georgian science and education.

\section{Introduction}\label{1s1}
This  paper is being revived after having been frozen, in an unfinished form, for more than a decade.  It is essentially an extended abstract of the author's 1998 dissertation \cite{diss}. For ``historical"  considerations, its original style and language are almost fully preserved. 

Since the birth of\index{Girard,J.Y.} Girard's \cite{Gir} \index{linear
logic\ \ \ }linear logic, the topic of \index{substructural logics\ \ \
}substructural logics, often
called ``resource logics", has 
drawn the attention of many researchers, with various motivations and
different traditions.\footnote{An extensive survey of substructural logics
can be found in \cite{Dosen}.}  
The common feature of these logics is that they are sensitive with 
respect to the number of occurrences of subformulas in a formula or a sequent, the most demonstrative
example of which is the failure of the 
classical principles \[A\rightarrow A\& A\] and 
\[\frac{A\rightarrow B \ \ \ \ \ \ A\rightarrow C}{A\rightarrow B\& C}\]
as a result of removing the rule of contraction from classical sequent calculus.

The philosophy behind this approach is that if formulas are viewed as resources, the conjunction
is viewed as an operator which ``adds up" resources and the implication is viewed as an operator
which converts one resource (the antecedent) into another (the consequent), then $A\& A$ 
is generally stronger than $A$, and $A\rightarrow(B\& C)$ is stronger than $(A\rightarrow
B)\& (A\rightarrow C)$. For example, $\$1\& \$1$ should be understood as being 
equivalent to $\$2$ rather than $\$1$, so that $\$1\rightarrow(\$1\&\$1)$ is not valid;
one cannot 
get both a can of Coke and a sandwich for $\$1$ even if each of them costs a dollar, so  
\[\frac{\$1\rightarrow coke \ \ \ \ \ 
\ \$1\rightarrow sandwich}{\$1\rightarrow(coke\ \&\  sandwich)}
\]
fails, too.

Although this kind of resource philosophy
seems intuitively very clear, natural and appealing, 
it has never been fully formalized, and substructural logics  
owe their name ``resource logics" mostly to certain syntactic features rather 
than some strict and, at the same time, intuitively convincing resource 
semantics behind them. The present work is an 
attempt to develop such a semantics.

\subsection{Resources, informally} 
 \index{resource}The simplest type of resources, which we call 
{\em unconditional resources},  consist of 2 components:  
{\em effect} and {\em potential}. \index{effect (of a resource)\ \ \ }
\index{potential (of a resource)\ \ \ }
Effect is the process associated with and supported by the resource, and potential is the 
set of resources into which the resource can be converted at the owner's wish.
The effect of the resource ``My car in my driveway" is the process ``my car is
in my 
driveway", and its potential, at a high level of abstraction, is \{``My car on
the Ross bridge", ``My car on the Franklin Bridge", ``My car at the airport",...\}.
For some resources, such as $2\times 2=4$, the potential can be empty. And some resources
can be ``effectless" --- their value is associated only with the  resources into 
which they can be converted. Money can be viewed as an example of 
such a resource: its value is associated only with
its convertibility into ``real" things.

The elements of the potential can be viewed as the {\em commands} that the resource accepts
from its owner. My computer, when it is shut down, maintains the process ``the screen 
is dark" as its effect. In this state it accepts only one command: ``start". After this command 
is given, it turns into another resource, --- let us say ``another resource" rather than ``another
state", --- whose effect is ``the initial menu is on the 
screen" and whose potential consists of the commands ``Word Processor", ``Games", ``Netscape",
``Telnet",... When I give one of these commands, it turns yet into a new resource, etc. 

It might be helpful to think of resources as {\em agents} carrying out 
certain {\em tasks}\index{task} for us. 
Mathematically, 
task is a synonym of resource, and which of these
two words we use, depends on the context. Between agents  a master-slave (ownership)
relationship can hold. What is a task for the slave, is a resource for the master.

Thus, intuitively, an unconditional resource is an agent which maintains a
certain process (its effect)
and also accepts commands (elements of its potential) from its master and executes them, where 
executing a command means turning into a certain new resource.

\subsection{Resource conjunction}
Let us consider some more precise examples. Let $\Phi$ be an agent  which writes in
memory location $L1$ any 
number
we tell it to 
write, and then keeps this number there until we give it a new command of the same 
type again, and so on. 
Initially, this resource maintains the value 0 in $L1$.

This is an example of an inexhaustible resource --- it can execute our commands as many 
times as we like.

Consider another agent $\Psi$ which writes in location $L2$, when we give it such a command, 
the factorial of the current value of $L1$, and keeps that number there
(even if the value of $L1$ changes meanwhile)
until we give it a new command of the same type.
 Unlike $\Phi$,
 this resource accepts only one command, even though, again, infinitely many times.
Initially it keeps the value 0 in $L2$.

We denote the conjunction operator for resources by $\rcc$. What would $\Psi\rcc\Phi$ mean?
Intuitively, having the conjunction
of two agents means having them both as independent resources, so that we can use 
each of them as we wish, without affecting our ability to use the other. 
Initially this resource maintains the value 0 in locations $L1$ and $L2$. In every state,
it accepts two 
types of commands: 1) Write in $L2$ and maintain there the factorial of the current value
of $L1$ (only one command), and 2) Write in $L1$ and maintain there number $n$ (one command
per number). 

\subsection{Conditional resources}
Both $\Psi$ and $\Phi$, as well as their conjunction $\Psi\rcc\Phi$, are examples of 
unconditional resources: they maintain their effects and execute commands unconditionally,
without asking anything in return.  However, in real life, most resources 
are {\em conditional}. My car can travel, but now and then it will require from me to 
fill up its tank; my computer will copy any file to a floppy disk, but only if I execute 
its countercommand ``Insert a floppy disk into drive A". 

We use the symbol $\rii$ to build expressions for conditional resources. 
Having the  resource $\Theta_1\rii\Theta_2$
means that I can make it work as $\Theta_2$ if I can give to it the
resource $\Theta_1$.
It is not necessary to actually assign some agent accomplishing
$\Theta_1$ to it. I can
assign to it a ``virtual $\Theta_1$", which means that all I need in order to make this
conditional resource 
work as $\Theta_2$ is to execute every command it issues for $\Theta_1$.
So,
$\Theta_1\rii\Theta_2$ can be seen as a resource which consumes the
resource $\Theta_1$
and produces the resource $\Theta_2$, or converts the resource
$\Theta_1$ into $\Theta_2$.

Consider one more unconditional resource, $\Gamma$, which writes in memory location $L2$ the  
factorial of any number we give to it, and maintains it there until it gets a new command
of the same type. Just like $\Psi$, initially it maintains 0 in $L2$.

Can I accomplish $\Gamma$ as a task? Generally --- not. Even if I can compute factorials in my 
head, I may not have writing access to location $L2$ after all, or I may have this 
access but some other agent can have that kind of access, too, and can overwrite the
number I needed to maintain in $L2$.

However, if the resources $\Phi$ and $\Psi$ are at my disposal, then 
I can carry out $\Gamma$. Whatever number my master gives me, I first make $\Phi$ 
write it in $L1$, and then make $\Psi$ write and maintain its factorial in 
$L2$. So, I cannot accomplish the task $\Gamma$, but I can accomplish the task
\[\Phi\rcc \Psi\ \rii\ \Gamma,\]
which is an example of a conditional resource.

If we go to lower levels of abstraction, it may turn out that, say,
$\Psi$, itself, is (the consequent of) a conditional
resource. It may require some memory space, ability to write and read
and perform some arithmetic operations there, etc. Let us denote this resource, --- the
resource required by $\Psi$ to function successfully, --- by
$\Delta$. In that case, the resource I possess is

\[\Phi\ \rcc\ (\Delta\rii\Psi)\]
rather than $\Phi\ \rcc\ \Psi$.
I have no reason to assume that I can  carry out $\Gamma$ now.
However, I can carry out
\[\Bigl(\Delta\ \rcc\ \Phi\ \rcc\ (\Delta\rii\Psi)\Bigr)\ \rii\
\Gamma.\]
Because I can use the conjunct $\Delta$ to do whatever $(\Delta\rii\Psi)$
wants from its $\Delta$, and thus make that conditional resource work as
$\Psi$.

What if $\Phi$, too, requires $\Delta$ as a resource? That is, can I successfully handle
the task
\[\Bigl(\Delta\ \rcc\ (\Delta\rii\Phi)\ \rcc (\Delta\rii\Psi)\Bigr)\
\rii\ \Gamma?\]

No, even though, by classical logic, the above formula follows from
\[\Bigl(\Delta\ \rcc\ \Phi\ \rcc\ (\Delta\rii\Psi)\Bigr)\ \rii\ \Gamma.\]

I cannot, because I have only one $\Delta$ while I need two. What if the two
conditional agents $\Delta\rii\Phi$ and $\Delta\rii\Psi$ issue
conflicting commands
for $\Delta$? For example, the first one may require to write in a certain location $L3$
the number 13 and
maintain it there, while the other needs 14 to be maintained in that location?
One location cannot keep two different values. In other words, $\Delta$ could serve
one master, but it may not be able to please two bosses simultaneously. And not only
because conflicting commands may occur. Certain resources can execute certain commands only
once or a limited number of times. A kamikaze can attack any target, and the commands 
``attack A" and ``attack B" are not logically conflicting; however, I cannot carry out 
the task of 2 kamikazes if I only have 1 kamikaze at my command: after making him 
attack A, he will be gone.

This is where linear-logic-like effects start to kick in.
As we have just observed, the
principle
$\Theta\ \rii\ \Theta\rcc\Theta$ is not valid. On the other hand, all
the principles of
linear logic + weakening are valid.

\subsection{More on our language}
In addition to $\rii$ and $\rcc$, we need many other
connectives to
make our language sufficiently expressive. When we described $\Phi$ and
$\Psi$, we
 just used English.
But our formal language should be able to express all that in formulas. In fact the
language is going to be much more
expressive than the language of linear logic.

The formulas of our language 
are divided into 3 categories/levels: facts, processes, and resources. Each
level has its own operators and may use lower-level expressions as subexpressions.

{\em Facts} are nothing but classical first order formulas, with their usual semantics.  We use the standard Boolean connectives and quantifiers
(without dots over them) to build complex facts from simpler ones. 

We assume that {\em time} is a linear order of {\em moments}, with a beginning but no end.
An {\em  interval} is given by a pair $(i,j)$, where $i$ is a time moment and $j$ is 
either a greater time moment or $\infty$.

While facts are true or false at time moments, {\em processes}
are true or false on intervals. 

The Boolean connectives and quantifiers are applicable to processes, too. To indicate
that they are process operators, we place one dot over them. $\alpha\pcc\beta$,
where $\alpha$ and $\beta$ are processes, is the process which is true on an interval
if and only if both $\alpha$ and $\beta$ are true on that interval. The meaning of
the other ``dotted" connectives ($\pii$, $\pnn$, $\puu$, ...) should also be clear. They behave just like classical
connectives, --- all the classical tautologies, with dots over the operators, hold
for processes, too. But, as we have already noted, this is not the case for resources.

\ 

Here are some other process operators: \label{proc1}

$\Uparrow A$, where $A$ is a fact, is a process
which holds on an interval iff $A$ is true at every moment of the interval except,
perhaps, its first moment.

$\angle A$, where $A$ is a fact, is true on an interval iff $A$ is true at the first 
moment of the interval.

$\alpha\rhd\beta$, where $\alpha$ and $\beta$ are resources,
 is true on an interval iff $\alpha$ holds on some initial segment of the
interval and $\beta$ holds on the rest of the interval; in other words, if the 
process $\alpha$ switches to the process $\beta$ at some internal moment of the interval.

As for the resource level expressions, they, too, use classical operators, with a double
dot over them. We have already seen the intuitive meaning of two of them, $\rcc$ and
$\rii$. 
The other basic resource-building operator is $\gg$. The expression 
\[\alpha\gg(\Delta_1,...,\Delta_n),\]
where $\alpha$ is a process and the $\Delta_i$ are resources, stands for the resource
whose effect is $\alpha$ and whose potential is $\{\Delta_1,...,\Delta_n\}$. The 
expression 
\[\alpha\gg x\Delta(x)\]
is used to express resources  with possibly infinite potentials: the potential of this resource 
is $\{\Delta(a):\ a\in D\}$, where $D$ is the domain over which the variable $x$ ranges.

To be able to express infinite resources such as $\Phi$ and $\Psi$ (Figures 1 and 2), we 
also need to allow recursively defined expressions. Let 

\[\Phi':=\ \ \gg x \Bigl(\bigl(\Uparrow L1(x)\bigr)\Phi'\Bigr)\]
 and
\[\Psi':=\ \ \gg\Bigl(\bigl(\pee x
(\angle(L1(x)\pcc\Uparrow L2(!x)\bigr)\Psi'\Bigr).\]

Then, resource $\Phi$ can be expressed by $\bigl(\Uparrow L1(0)\bigr)\Phi'$ and resource 
$\Psi$ can be expressed by $\bigl(\Uparrow L2(0)\bigr)\Psi'$.

For readers familiar with linear logic, we will note that $\gg$ is in fact a generalization
of the additive conjunction or quantifier (while $\rcc$ and $\rii$ correspond to the multiplicative
conjunction and implication). The generalization consists in adding one more parameter, $\alpha$, to this
sort of conjunction. The standard linear-logic additive conjunction should be viewed 
as a special case of $\gg$-formulas where the left argument of $\gg$  is a trivial process,
such as $\Uparrow \ftt$. 

The semantics of $\gg$ is that it is a ``manageable $\rhd$". If in
$\alpha\rhd\beta$ the
transfer from $\alpha$ to $\beta$ happens ``by itself" at an arbitrary
moment, in the case
of $\alpha\gg(\Psi_1,...,\Psi_n)$ the transfer from $\alpha$
to the effect of $\Psi_i$ happens according to our
command. But at what moment should this transfer occur? If we assume that exactly at 
the moment of giving the command, then even the principle $\Delta\rii\Delta$ can fail,
because execution of a command, or passing the command which I receive in the right
$\Delta$ to the left $\Delta$ always takes time. Hence, 
we assume the following protocol for 
$\alpha\gg(\Psi_1,...,\Psi_n)$: at some time moment $t$ and some $1\leq i\leq n$,
master decides to issue the command 
\[ DO(\Psi_i).\]
Then, at some later moment $t'$, slave is expected to explicitly report an execution of this 
command:
\[DONE(\Psi_i).\]
The transfer from $\alpha$ to the effect of $\Phi_i$ is assumed to take place at some 
moment between $t$ and $t'$.

For potential real-time applications,\label{1realtime} we may want to introduce a deadline parameter for $\gg$:
\[\alpha\gg^t(\Phi_1,...,\Phi_n).\]

This means that at most time $t$ should elapse between ``DO" and
``DONE". Another operator for which we might want to introduce a real-time 
parameter is $\rhd$: \ \ $\alpha\rhd^t\beta$ is a process which is true on an interval
$(i,j)$ iff there is $e$ with $i<e\leq i+t<j$ such that $\alpha$ is true 
on $(i,e)$ and $\beta$ is true on $(e,j)$. We leave exploring this possibility 
for the future, and the formal definitions of our language and semantics 
the reader will find in the later sections deal only with the non-real-time 
version.\footnote{The reader will also find that the language which we described here
is a slightly simplified version of our real formalism.}

\subsection{Our logic as a planning logic}

Later sections contain examples showing how our logic can be used as a 
 planning logic.
 A planning problem is represented as $\Delta\rii\Gamma$, where 
$\Gamma$ is a specification of the goal as a task, and $\Delta$ is the conjunction of the 
resources  we possess.
An action is understood as giving a command to one of these resources or, --- at a higher level 
of abstraction, ---  assigning one resource to another, conditional, resource. 
Hence, actions change only those (sub)resources to which they are applied.  The  
effect of an action for the rest of the resources is ``no change", and it is this property
that makes the logic frame-problem-free. Some examples also show how our logic can naturally handle
certain planning problems which, under the traditional approach, would require special 
means for representing knowledge.

\section{Facts}\label{1s2}

The components of our language, shared by all three types of expressions (facts, processes
and resources), are {\em variables} and {\em constants}. The set of variables is infinite.
The set of constants may be finite or infinite. For now, we will make a simplifying 
assumption that the set of constants is $\{0,1,2,...\}$. The set of {\em terms} is the union 
of the set of variables and the set of constants.

  We also have a 
set of {\em fact letters} (called predicate letters in classical logic), with each of which
is associated a natural number called {\em arity}.

\

{\em Facts}\index{fact} are the elements of the smallest set $F$ of expressions, such that, saying
``$A$ is a fact" for ``$A\in F$", we have:

 \begin{itemize}
\item $\fff$ is a fact;
\item if $P$ is an $n$-ary fact letter and $t_1,...,t_n$ are terms, then $P(t_1,...,t_n)$
is a fact;
\item if $A$ and $B$ are facts, then $(A)\fii(B)$ is a fact;
\item if $A$ is a fact and $x$ is a variable, then $\fuu x(A)$ is a fact. 
\end{itemize}

As we see, facts are nothing but formulas of classical first order logic. In the sequel,
we will often omit some parentheses when we believe that this does not lead to ambiguity.

The other classical operators are introduced as standard abbreviations:
\begin{itemize}
\item $\fnn A=A\fii \fff$;
\item $A\fdd B=(\fnn A)\fii B$;
\item $A\fcc B=\fnn(\fnn A\fdd\fnn B)$;
\item $A \feq B=(A\fii B)\fcc(B\fii A)$;
\item $\ftt=\fnn\fff$;
\item $\fee xA=\fnn\fuu x\fnn A$.
\end{itemize}

A {\em free variable} of a fact is a variable $x$ which has an occurrence in the fact which
is not in the scope of $\fuu x$ (not {\em bound} by $\fuu$).
A fact is {\em closed}, if it has no free variables.

A {\em situation} is a set $s$ of closed facts such that, using the expression $s\models A$ 
for $A\in s$, we have:\begin{itemize}
\item $s\not\models \fff$;
\item $s\models A\fii B$ iff $s\not\models A$ or $s\models B$;
\item $s\models \fuu xA(x)$ iff $s\models A(a)$ for every  constant $a$.
\end{itemize}

If $s\models A$, we say that $A$ is {\em true}, or {\em holds} in  situation $s$.

\ 

We fix an infinite set ${\cal T}$ of {\em time moments} together with a strict linear 
ordering relation
$<$ on it. $i\leq j$, as one can expect, means $i<j$ {\em or} $i=j$. We assume that 
$0\in{\cal T}$ and, for all $i\in{\cal T}$, \ $0\leq i$.

${\cal T}^+$ denotes 
the set ${\cal T}\cup\{\infty\}$. The ordering relation $<$ is extended to ${\cal T}^+$ 
by assuming that for all $t\in {\cal T}$, \ $t<\infty$.

An {\em interval} is a pair $(i,j)$, where $i\in{\cal T}$, $j\in{\cal T}^+$ and $i<j$.

A {\em world} is a function $W$ which assigns to every time moment $i\in{\cal T}$ a 
situation $W(i)$.

\section{Processes}\label{1s3}

\index{process}This section contains formal definitions for the syntax and the semantics
of the process level of our language. The reader can be advised to go
back to page \pageref{proc1} to refresh their memory regarding the
intuition and the motivation behind the key process operators introduced 
below. Some later sections --- first of all, Section \ref{1s7}, --- contain
examples that could also help the reader to better understand our
formalism in action.

The same applies to the next section, where we give  formal
definitions for the resource level, the motivation and intuition being
explained in the introductory section.

\begin{definition}
\label{1d1}{\em The set of \index{finitary process}{\em finitary processes} is the smallest set $FP$ of 
expressions such that, saying ``$\alpha$ is a finitary process" for ``$\alpha\in FP$",
we have:

\begin{enumerate}
\item if  $A$ is a fact, then $\angle A$ is a finitary process;
\item if $A$ is a fact, then $\updownarrow A$ is a finitary process;
\item if $A$ is a fact, then $\Uparrow A$ is a finitary process;
\item if $A$ is a fact, then $\Box A$ is a finitary process;
\item if $\alpha$ and $\beta$ are finitary processes, then so
is $\alpha\pii\beta$;
\item if $\alpha$ is a finitary process and $x$ is a variable, then $\puu x\alpha$
is a finitary process;
\item if $\alpha$ and $\beta$ are finitary processes, then so is $\alpha\rhd\beta$;
\item if $\alpha$ is a finitary process, then so is $[\rhd]\alpha$;
\item if $\alpha$ is a finitary process, then so is $[\unrhd]\alpha$.
\end{enumerate}
} \end{definition}

Some other process operators are introduced as abbreviations:
\begin{itemize}
\item $\pff=\Box\fff$;
\item $\ptt=\angle\ftt$;
\item $\pnn \alpha=\alpha\pii \pff$;
\item $\alpha\pdd \beta=(\pnn \alpha)\pii \beta$;
\item $\alpha\pcc \beta=\pnn(\pnn \alpha\pdd\pnn \beta)$;
\item $\alpha \peq \beta=(\alpha\pii \beta)\pcc(\beta\pii \alpha)$;
\item $\pee x\alpha=\pnn\puu x\pnn \alpha$;
\item $\Updownarrow\alpha=\angle\alpha\pcc\Uparrow\alpha$;
\item $\Downarrow\alpha=\angle\alpha\pcc\updownarrow\alpha$;
\item $\alpha\unrhd\beta=\alpha\pdd(\alpha\rhd\beta)$.
\end{itemize}

A {\em closed process} is a process in which every variable is bound by $\fuu$ or $\puu$.

\begin{definition}
\label{1d2}{\em {\em Truth} of a closed finitary process $\gamma$ on an interval $(i,j)$
in a world $W$, symbolically $W\models_{i,j}\gamma$, is defined as follows:
\begin{itemize}
\item  $W\models_{i,j}\angle A$ iff $W(i)\models A$;
\item  $W\models_{i,j}\updownarrow A$ iff for all $r\in\cal{T}$ with $i<r< j$, \ 
 $W(r)\models A$;
\item  $W\models_{i,j}\Uparrow A$ iff for all $r\in\cal{T}$ with $i<r\leq j$, \
 $W(r)\models A$;
\item  $W\models_{i,j}\Box A$ iff for all $r\in\cal{T}$,  \
 $W(r)\models A$;
\item $W\models_{i,j}\alpha\pii\beta$ iff $W\not\models_{i,j}\alpha$ or 
$W\models_{i,j}\beta$;
\item $W\models_{i,j}\puu x\alpha(x)$ iff for every constant $a$,
\ $W\models_{i,j}A(a)$;
\item $W\models_{i,j}\alpha\rhd\beta$ iff there is $e$ with $i<e<j$ such that $W\models_{i,e}
\alpha$ and $W\models_{e,j}\beta$;
\item $W\models_{i,j}[\rhd]\alpha$ iff there are $e_0,e_1,e_2,...\in{\cal T}$ with $e_0=i$, $e_0<e_1<e_2<...<
j$ such that for every $k$, $W\models_{e_k,e_{k+1}}\alpha$;
\item $W\models_{i,j}[\unrhd]\alpha$ iff $W\models_{i,j}[\rhd]\alpha$ or there are $e_0,
...,e_n\in{\cal T}^+$ such that $e_0=i$, $e_n=j$ and, for every $k:\ 0\leq k<n$, 
$W\models_{e_k,e_{k+1}}\alpha$.
\end{itemize} }\end{definition}

When we later define a semantics for resources, we will also have to deal with 
 infinite process expressions. 

\begin{definition}
\label{1d3}{\em An {\em infinitary process}\index{infinitary process} is defined by replacing in the definition
of finitary process the word ``finitary" by ``infinitary" and adding the following 3 clauses:
\begin{itemize}
\item if $\alpha_0,\alpha_1,\alpha_2,...$ are infinitary processes, then so is 
$\alpha_0\pcc\alpha_1\pcc\alpha_2\pcc...$;
\item if $\alpha_0,\alpha_1,\alpha_2,...$ are infinitary processes and $k_1,{k'}_1,k_2,{k'}_2,...,
\in {\cal T}$, $k_1<{k'}_1<k_2<{k'}_2<...$, then 
$\alpha_0\leadsto^{{k'}_1}_{k_1}\alpha_1\leadsto^{{k'}_2}_{k_2}\alpha_2...$ is an 
infinitary process;
\item if $\alpha_0,...,\alpha_n$ ($n\geq 1)$ are infinitary processes and $k_1,{k'}_1,...,
k_n,{k'}_n\in {\cal T}$, $k_1<{k'}_1<...<k_n<{k'}_n$, then 
$\alpha_0\leadsto^{{k'}_1}_{k_1}...\leadsto^{{k'}_n}_{k_n}\alpha_n$ is an 
infinitary process.
\end{itemize} }\end{definition}

Thus, every finitary process is, at the same time, an infinitary process, but not 
vice versa. We will use the common name ``process" for both types of processes.

\begin{definition}
\label{1d4}{\em To get the definition of {\em truth} for closed infinitary processes, we replace
the word ``finitary" by ``infinitary" in Definition \ref{1d2} and add the following 3 clauses:
\begin{itemize}
\item $W\models_{i,j}\alpha_0\pcc\alpha_1\pcc\alpha_2\pcc...$ iff for every $k$, \ 
$W\models_{i,j}\alpha_k$;
\item $W\models_{i,j}\alpha_0\leadsto^{{k'}_1}_{k_1}\alpha_1\leadsto^{{k'}_2}_{k_2}\alpha_2...$
iff there are $e_0,e_1,e_2,...\in{\cal T}$ such that
$e_0=i$ and, for each $m\geq 1$ we have 
$k_m<e_m<{k'}_m$ and $W\models_{e_{m-1},e_m}\alpha_{m-1}$.
\item $W\models_{i,j}\alpha_0\leadsto^{{k'}_1}_{k_1}...\leadsto^{{k'}_n}_{k_n}\alpha_n$
iff there are $e_0,,...e_{n+1}\in{\cal T}$ such that $e_0=i$,
$e_{n+1}=j$, for each $m: \ 1\leq 
m \leq n$ we have \ $k_m<e_m<{k'}_m$ and, for each $m: \ 1\leq m\leq n+1$ we 
have  $W\models_{e_{m-1},e_m}\alpha_{m-1}$.
\end{itemize} }\end{definition}

\begin{definition}\label{1d5}{\em
A closed process is said to be {\em valid} iff it is true on every interval of every world.}
\end{definition}

Here are some simple observations (all processes are assumed to be closed):

All processes of the form of a classical tautology, such as, say, $\alpha\pdd\pnn\alpha$
or $\alpha\pii\alpha\pcc\alpha$,  are valid.

$(\Uparrow A\pcc\Uparrow B)\ \peq\ \Uparrow(A\fcc B)$ is valid.

$(\angle A\pdd\angle B)\ \peq\ \angle(A\fdd B)$ is valid.

$(\Uparrow A\pdd \Uparrow B)\ \pii\ \Uparrow(A\fdd B)$ is valid, but $\Uparrow(A\fdd B)\ \pii\ 
(\Uparrow A\pdd\Uparrow B)$ is not.

$\rhd$ is associative: $\bigl(\alpha\rhd(\beta\rhd\gamma)\bigr)\peq\bigl((\alpha\rhd\beta)
\rhd\gamma\bigr)$ is valid.

\section{Resources}\label{1s4}

At this level of the language 
we have 3 sorts of expressions: resources, DO-resources and DONE-resources. 
We define them simultaneously:

\begin{definition}\label{1d6}{\em \ \ 

\begin{enumerate}
\item A \index{DO-resource}{\em DO-resource} is an expression $\gg x_1,...,x_m(\Phi_1,...,\Phi_n)$ ($m\geq 0,\ 
n\geq 0$), where the $x_i$ are variables and the $\Phi_i$ are DONE-resources. If $n=1$, we can 
omit the parentheses and write just $\gg x_1,...,x_m\Phi$; if $n=0$, we write just $\gg$.
\item A {\em \index{DONE-resource}DONE-resource} is an expression $\ll x_1,...,x_m(\Phi_1,...,\Phi_n)$ ($m\geq 0,\ 
n\geq 1$), where the $x_i$ are variables and the $\Phi_i$ are resources. If $n=1$, we can 
omit the parentheses and write just $\ll x_1,...,x_m\Phi$; 
\item A {\em resource}\index{resource} is one of the following:\begin{itemize}
\item $\alpha\Phi$, where $\alpha$ is a finitary process 
and $\Phi$ is a DO-resource;
\item $\Phi\rii\Psi$, where $\Phi,\Psi$ are resources;
\item $\Phi\rcc\Psi$, where $\Phi,\Psi$ are resources;\footnote{In fact, $\Phi\rcc\Psi$
can be defined as $\rnn(\rnn\Phi\pdd\rnn\Psi)$ (see the $\rcc$-independent definition of $\rnn$ and $\rdd$
below), but we still prefer to treat $\rcc$ as a basic symbol.}
\item $\ruu x\Phi$,  where $x$ is a variable and $\Phi$ is a resource.
\end{itemize}
\end{enumerate}
 }\end{definition}
We also introduce  the following abbreviations:

\begin{itemize}

\item $\rff=\pff\gg$;
\item $\rtt=\ptt\gg$;
\item $\rnn\Phi=\Phi\rii \rff$;
\item $\Phi\rdd\Psi=(\rnn\Phi)\rii\Psi$;
\item $\ree x\Phi=\rnn\ruu x\rnn\Phi$;
\item $\Phi\acc\Psi=\ptt\gg(\ll \Phi, \ll \Psi)$;
\item $\auu x\Phi= \ptt\gg x\ll\Phi$;
\item $\Phi\add\Psi=\rnn(\rnn\Phi\acc\rnn\Psi)$;
\item $\aee x\Phi=\rnn(\auu x\rnn\Phi)$.
\end{itemize}

Thus, every resource is a $(\rii,\rcc,\ruu)$-combination of expressions of the type
\[\alpha\gg \vec{x}\Bigl(\ll\vec{y_1}\bigl(\Phi_{1}^{1}(\vec{x},\vec{y_1}),
\ldots,\Phi_{k_1}^{1}(\vec{x},\vec{y_1})\bigr) ,\ldots, \ll\vec{y_n}\bigl(\Phi_{1}^{n}(\vec{x},\vec{y_n}),
\ldots,\Phi_{k_n}^{n}(\vec{x},\vec{y_n})\bigr)\Bigr) ,\]
where $\alpha$ is a finitary process and the $\Phi_{i}^{j}$ are resources. This expression 
represents an agent
which maintains the process $\alpha$ as its effect; a command to it should be given by specifying $\vec{x}$
as a sequence $\vec{a}$ of constants, and specifying one of the $i$, \ $1\leq i\leq n$. 
The expression to the right of $\gg$ represents the potential of this resource.
We see that this potential is 
more complex than the type of potentials discussed in Section \ref{1s1}. The intuitive 
meaning of $\ll\vec{y_i}\bigl(\Phi_{1}^{i}(\vec{a},\vec{y_i}),
\ldots,\Phi_{k_i}^{i}(\vec{a},\vec{y_i})\bigr) $ as a command is that slave has to 
produce the resource $\Phi_{j}^{i}(\vec{a},\vec{b})$ for a $j$ ($1\leq j\leq k_i$) and $\vec{b}$ of his own choice, 
and report about this choice (along with the fact of executing the command) to master.

\ 

The operators $\gg$ and $\ll$, when followed by a nonempty list of variables, act as 
quantifiers: they bind the occurrences of these variables within their scope.
An occurrence of a variable in a resource is said
to be {\em free}, if it is not in the scope of $\gg$, $\ll$, $\fuu$, $\puu$ or $\ruu$.
If a resource does not contain free occurrences of 
variables, then it is said to be {\em closed}.

Note that our definition of resource allows infinite expressions: there is no 
requirement ``... is the smallest set of expressions such that...". Naturally, we only
want to deal with resources which can be expressed finitarily. One way to express 
a certain class of infinite resources finitarily is to allow {\em recursive definitions}
for their subexpressions. For safety, we will only allow definitions that
have the following form:\label{1recdef}
\begin{equation}\label{1e100}
\Phi:=\ \ \gg \vec{x}\Bigl(\ll\vec{y_1}(\alpha_{1}^{1}\Phi_{1}^{1},
\ldots,\alpha_{k_1}^{1}\Phi_{k_1}^{1}) ,\ldots, \ll\vec{y_n}(\alpha_{1}^{n}\Phi_{1}^{n},
\ldots,\alpha_{k_n}^{n}\Phi_{k_n}^{n})\Bigr) ,\end{equation}
where the $\alpha_{i}^{j}$ are finitary processes and the $\Phi_{i}^{j}$ are DO-resources
 introduced by the same type (\ref{1e100}) of recursive definitions, so that, $\Phi$,
itself, can be among the $\Phi_{i}^{j}$. 

A recursive definition is not a part of a formula but should be
given separately, and if a resource contains a recursively defined subexpression $\Phi$, we assume that 
$\Phi$ just abbreviates the right-hand side of its definition.

Another type of finitarily represented infinite expressions we will allow in resources
is \[!\Phi,\]
which is understood as an abbreviation for the infinite conjunction
\[\Phi\rcc\Phi\rcc\Phi\rcc...\]

We will call the resources that are finite expressions, possibly containing $!$-expressions
and recursively defined 
subexpressions of the form (\ref{1e100}), {\em finitary resources}. Since we are going to deal  only with this type of 
resources, from now on, the word ``resource" will always refer to finitary resources.

\  

We are now going to give a formal definition of the semantics for resources. This 
definition is in a game-semantical style as we understand a resource  as a potential
game between master and slave, where moves consist in giving commands and/or
reporting execution of commands. 
 
\  

A {\em position}\index{position} is one of the following:\begin{itemize}
\item a  resource;
\item a  DONE-resource;
\item $\Phi\rii\Psi$, where $\Phi$ and $\Psi$ are positions;
\item $\Phi\rcc\Psi$, where $\Phi$ and $\Psi$ are positions;
\item $\ruu x\Phi$, where $x$ is a variable and $\Phi$ is a position.
\end{itemize}

When speaking about a subexpression of an expression, we are often interested in a concrete
occurrence of this subexpression rather than the subexpression as an expression (which may have several
occurrences). In order to stress that we mean a concrete occurrence, we shall use the words 
\index{osubexpression}``osubexpression", \index{osubposition}``osubposition", etc. (``o" for ``occurrence"). 

\ 

A {\em surface osubexpression} of a resource or a
position is an osubexpression which is not in the scope of $\gg$ or $\ll$. 

Such an osubexpression is {\em positive}, or has a {\em positive occurrence}, if it is in the 
scope of an even number of $\rii$; 
otherwise it is {\em negative}.

\begin{definition}\label{1d7}{\em 
A \index{master's move}{\em master's move} for a position $\Phi$ is a position which results from 
$\Phi$ by \begin{itemize}
\item  replacing some finite (possibly zero) number of positive surface osubpositions of the form 
$\alpha\gg\vec{x}\bigl(\Psi_1(\vec{x}),...,\Psi_n(\vec{x})\bigr)$ by $\Psi_i(\vec{a})$ for some sequence
$\vec{a}$ (of the same length as $\vec{x}$) of constants and some  $i$ 
($1\leq i\leq n$), and/or 
\item replacing some finite (possibly zero) number of negative surface osubpositions of the form
$\ll\vec{x}\bigl(\Psi_1(\vec{x}),...,\Psi_n(\vec{x})\bigr)$ by $\Psi_i(\vec{a})$ for
some $\vec{a}$ and $i$ ($1\leq i\leq n$).
\end{itemize}

\index{slave's move}{\em Slave's move} is defined in the same way, only with 
the words ``positive" and ``negative" interchanged.}\end{definition}  

Thus, master's moves consist in giving commands in positive osubresources and reporting execution
of commands in negative osubresources, while
slave's moves consist in giving commands in negative osubresources and reporting execution
of commands in positive osubresources.

Suppose $\Psi'$ and $\Psi''$ are master's and slave's moves for a position $\Phi$. Then
the {\em composition} of these two moves with respect to $\Phi$ is the position $\Psi$ which results from 
$\Phi$ by combining all the changes made by master and slave in $\Phi$ in their $\Phi$-to-$\Psi'$ and
$\Phi$-to-$\Psi''$ moves. $\Psi$ is said to be a 
{\em move} for $\Phi$. Note  that every position is a move for itself.

\

For a position $\Phi$, a {\em $\Phi$-play},\index{play\ \ \} or a {\em play over $\Phi$},  is a
finite or infinite sequence of the type 
\[\langle \Phi_0,t_1,\Phi_1,t_2,\Phi_2,...\rangle,\]
where  $\Phi_0=\Phi$, the $t_i$ are increasing  time moments ($t_1<t_2<...$) and, for each 
$i$,  $\Phi_{i+1}$ is a move for $\Phi_i$. 

 A play is said to be {\em compact},\index{compact play\ \ \ } if no two
neighboring positions ($\Phi_i$ and $\Phi_{i+1}$) in it are identical. If a play $P$ is not 
compact,
its {\em compactization}, denoted by $P^+$, 
is the play which results from $P$ by deleting every position which is identical
to the previous position, together with 
the time moment separating these two positions.

\

Intuitively, the $\Phi_i$ are the consecutive positions of the play, and $t_i$ is the 
time moment at which the position $\Phi_{i-1}$ is replaced by $\Phi_i$.

\ 

Note that the $(\rii,\rcc,\ruu)$-structure of a position in a play is inherited by the subsequent 
positions. 

Every (compact) play $P$  produces a unique process $P^*$ defined below. In this definition, ``..." 
does not necessarily mean an infinite continuation of the list (play): such a list can be 1-element,
$n$-element or infinite; in clause 6, $\vec{Q}$ stands for an arbitrary (possibly empty) sequence 
$t_1,\Gamma_1,t_2,\Gamma_2,...$. 

\begin{definition}\label{1d8}{\em (of the operation $\ ^*$)
\begin{enumerate}

\item  \(\langle \alpha\gg\Phi\rangle^*\ =\ \alpha.\)

\item \(\langle \Phi_0\rii\Psi_0,\  t_1,\ 
\Phi_1\rii\Psi_1, \ t_2,\ ...\rangle^*\ =\\
\langle\Phi_0,\ t_1,\ \Phi_1,\ t_2,\ ...\rangle^{+*}\pii 
\langle\Psi_0, \ t_1,\ \Psi_1,\  t_2,\ ...\rangle^{+*}.\)

\item \(\langle \Phi_0\rcc\Psi_0,\  t_1,\ 
\Phi_1\rcc\Psi_1, \ t_2,\ ...\rangle^*\ =\\
\langle\Phi_0,\ t_1,\ \Phi_1,\ t_2,\ ...\rangle^{+*}\pcc 
\langle\Psi_0, \ t_1,\ \Psi_1,\  t_2,\ ...\rangle^{+*}.\)

\item \(\langle \ruu x\Phi_0,\ t_1,\ \ruu x\Phi_1,\ t_2,\ ...\rangle^*\ =
\puu x(\langle\Phi_0,\ t_1,\ \Phi_1, \ t_2,\ ..\rangle^*).\)

\item  $\langle\alpha\gg\vec{x}(\Psi_1(\vec{x}),...,\Psi_n(\vec{x})), t,
\Psi_i(\vec{a})\rangle^*= \pff$.

\item  If \(P=\langle\\
\begin{array}{c}
\alpha\gg\vec{x}\Bigl(
\ll\vec{y_1}\bigl(\Psi_{1}^{1}(\vec{x},\vec{y_1}),...,\Psi_{k_1}^{1}(\vec{x},\vec{y_1})\bigr),
...,
\ll\vec{y_n}\bigl(\Psi_{1}^{n}(\vec{x},\vec{y_n}),...,\Psi_{k_n}^{n}
(\vec{x},\vec{y_n})\bigr)\Bigr), \\
k, \\
\ll\vec{y_i}\bigl(\Psi_{1}^{i}(\vec{a},\vec{y_i}),...,\Psi_{k_i}^{i}(\vec{a},\vec{y_i})\bigr),\\
m,\\
\Psi^{i}_{j}(\vec{a},\vec{b}),\\
\vec{Q}\rangle,

\end{array}
\)

then \[P^*\ =\  \alpha\leadsto^{m}_{k}\langle\Psi^{i}_{j}(\vec{a},\vec{b}),
\vec{Q}\rangle^*.\]

\end{enumerate}
}\end{definition}

\ 

{\bf Explanation:} According to clause 3, a play over a $\rcc$-conjunction of
resources produces the $\pcc$-conjunction of the processes produced
by the (sub)plays over the conjuncts of the resource. Similarly for 
the other double-dotted connectives $\rii$ (clause 2) and $\ruu$ 
(clause 4). 

A play over 
\[\alpha\gg\vec{x}\Bigl(
\ll\vec{y_1}\bigl(\Psi_{1}^{1}(\vec{x},\vec{y_1}),...,\Psi_{k_1}^{1}(\vec{x},\vec{y_1})\bigr),
...,
\ll\vec{y_n}\bigl(\Psi_{1}^{n}(\vec{x},\vec{y_n}),...,\Psi_{k_n}^{n}
(\vec{x},\vec{y_n})\bigr)\Bigr)
\]
produces $\alpha$, if no moves have been made (clause 1). If a command
\[\ll\vec{y_i}\bigl(\Psi_{1}^{i}(\vec{a},\vec{y_i}),...,\Psi_{k_i}^{i}
(\vec{a},\vec{y_i})\bigr) \]
was given but a report never followed (clause 5), 
we consider this  a failure of the non-reporting resource to carry out its task,
and associate the always-false process $\pff$ with this play so that it is never successful.
Finally, if a report 
$\Psi_{j}^{i}(\vec{a},\vec{b})$ followed the command, the play produces $\alpha\leadsto^{m}_{k}\beta$,
where $k$ is the moment of giving the command, $m$ is the moment of reporting 
its execution, and $\beta$ is the process
produced by the subplay over $\Psi_{j}^{i}(\vec{a},\vec{b})$; truth of this $\leadsto$-process means 
that the process $\alpha$ switches to the process $\beta$ at some time after the command and before the 
report.

One can show that as long as $\Phi$ is a closed finitary process, the process $P^*$
produced by a $\Phi$-play $P$ is always a closed infinitary process in the sense of 
Definition \ref{1d3}.

\

A {\em slave's strategy} is a function $f$ which assigns to every position 
$\Phi$ a slave's move for $\Phi$. We assume 
that this function is implemented as a program on a machine, and we
denote by $f'(\Phi)$ 
the time this program takes to give an output for input $\Phi$; if the program doesn't 
give any output, or gives an output which is not a slave's move for
$\Phi$, then we assume that $f'(\Phi)=\infty$.

\ 

Let $\Phi_0$ be a resource and $f$ be a slave's strategy. Here is 
an informal definition of  a {\em $\Phi_0$-play with
slave's strategy $f$}. The play starts at moment 0, and at this stage it is the one-position (sub)play 
$\langle \Phi_0\rangle$. Slave, i.e. the function $f$, takes $\Phi_0$ as an input, and starts computing
an output for it, --- thinking what move to make for $\Phi_0$. While slave is thinking, master 
can make some moves $\Phi_1,...,\Phi_{n}$ at time moments $t_1,...,t_{n}$, where $n\geq 0$, 
$t_1<...<t_n$ and each $\Phi_i$ ($1\leq i\leq n$) is a master's move for $\Phi_{i-1}$. Note that  
$\Phi_n$ is a master's move for $\Phi_0$ by the transitivity of this relation. 
The play has thus evolved to \[\langle \Phi_0, t_1,\Phi_1,...,t_{n},\Phi_{n}\rangle.\] 

Finally, at moment $t_{n +1}=f'(\Phi_0)$, $f$ computes a slave's move $\Psi$ for $\Phi_0$, and 
the next two items of the play become $t_{n +1}$ and $\Phi_{{n}+1}$, where $\Phi_{{n}+1}$ is 
the composition of $\Psi$ and $\Phi_{n}$ with respect to $\Phi_0$. Note that $\Phi_{n+1}$ is, at 
the same time, a slave's move for $\Phi_n$. 

So far slave has been busy processing 
the input $\Phi_0$ and did not see master's moves. Only now he looks at the current (last) position and 
sees that it is $\Phi_{{n}+1}$. So, he takes this position as a new input, and starts computing 
a move for it. While slave is thinking on his second move, master can continue making moves and 
the play can evolve to
\[\langle \Phi_0, t_1,\Phi_1,...,t_{n},\Phi_{n},t_{n +1},\Phi_{n +1},...,t_{m},\Phi_{m}\rangle\]
until, at some moment $t_{{m}+1}$, slave 
comes up with a move $\Gamma$ for $\Phi_{{n}+1}$.   The next two items of the 
play become  $t_{{m}+1}$ and  $\Phi_{{m}+1}$, where $\Phi_{{m}+1}$ is the composition of $\Gamma$ and 
$\Phi_{m}$ with respect to $\Phi_{{n}+1}$. And so on...

If, at some stage, $f$ fails to compute a move, that is, thinks for an infinitely long time,
 then all the further
moves will be made only by master. In this case, master may make not only a finite, but also an infinite number
of consecutive moves.

\

 We say that a play $P$ is {\em successful} with respect to a world $W$, iff
$W\models_{0,\infty} P^*$.

 A slave's strategy is said to be {\em universally successful}\index{universally successful strategy\ \ \ } for 
a closed resource $\Phi$, iff every $\Phi$-play with this strategy is successful with respect
to every world.

\begin{definition}\label{1d9}{\em 
We say that a resource  is {\em universally valid}\index{universally
valid resource\ \ \ }
iff there is a universally successful slave's strategy for it.}
\end{definition}

\section{Resource schemata}\label{1s5}

In this section we extend our language by adding to it {\em resource letters}. Formulas
of this extended language can be viewed as schemata for resources, where resource 
letters stand for resources.

Every resource letter has a fixed arity. The definition of 
\index{resource scheme\ \ \ }{\em resource scheme} is the 
same as the definition of resource (where the word ``resource" is replaced by 
``resource scheme"), with the following additional clause:
\begin{itemize}
\item if $\Phi$ is an $n$-ary resource letter and $t_1,...,t_n$ are terms, then 
$\Phi(t_1,...,t_n)$ is a resource scheme.\end{itemize}

For safety, we assume that the set of variables occurring in resource schemata is 
a proper subset of the set of variables of the language introduced in the previous 
sections, and that there are infinitely many variables in the latter that don't
occur in resource schemata.

\

A resource is said to be {\em safe} if it is $\ptt\Phi$ for some DO-resource
$\Phi$, or a ($\rii,\rcc,\ruu$)-combination of resources of this
type. 

Safe resources are what we could call ``effectless" resources: they 
are not responsible for maintaining any nontrivial process and their 
value is associated only with their convertibility into other
resources.

\ 

 A {\em substitution}\index{substitution\ \ \ }\index{safe
substitution\ \ \ } (resp. {\em safe substitution}) is a function $\tau$ which assigns to every 
$n$-ary resource letter $\Phi$  a resource (resp. safe resource) $\tau \Phi=\Psi(x_1,...,x_n)$ with exactly
$n$ free variables which does not contain any variables that might occur in resource
schemata.

Given a resource scheme $\Phi$ and a substitution $\tau$, $\Phi^{\tau}$ is defined as 
a result of substituting in $\Phi$ every resource letter $P$ by $\tau P$. More precisely,
\begin{itemize}
\item for an atomic resource scheme $\Phi$ of the form $P(t_1,...,t_n)$, where the $t_i$ are terms and
 $\tau P=\Psi(x_1,...,x_n)$, we have 
$\Phi^{\tau}=\Psi(t_1,...,t_n)$;
\item $\Bigl(\alpha\gg \vec{x}\Bigl(\ll\vec{y_1}(\Phi_{1}^{1},
\ldots,\Phi_{k_1}^{1}) ,\ldots, \ll\vec{y_n}(\Phi_{1}^{n},
\ldots,\Phi_{k_n}^{n})\Bigr)\Bigr)^{\tau}=\\
\alpha\gg \vec{x}\Bigl(\ll\vec{y_1}((\Phi_{1}^{1})^{\tau},
\ldots,(\Phi_{k_1}^{1})^{\tau}) ,\ldots, \ll\vec{y_n}((\Phi_{1}^{n})^{\tau},
\ldots,(\Phi_{k_n}^{n})^{\tau})\Bigr)$;  
\item $(\Phi\rii\Psi)^{\tau}=\Phi^{\tau}\rii\Psi^{\tau}$;
\item $(\Phi\rcc\Psi)^{\tau}=\Phi^{\tau}\rcc\Psi^{\tau}$;
\item $(\ruu x\Phi)^{\tau}=\ruu x(\Phi^{\tau})$.
\end{itemize}

We say that a resource $\Phi$ is an {\em instance}\index{instance (of a
resource)\ \ \ } of a resource scheme $\Psi$, iff
$\Phi=\Psi^{\tau}$ for some substitution $\tau$. If $\tau$ is a safe substitution, then
$\Phi$ is said to be a {\em safe instance}\index{safe instance\ \ \ } of $\Psi$.

\begin{definition}\label{1d10}{\em 
We say that a resource scheme $\Psi$ is {\em universally valid} (resp. {\em universally
s-valid})\index{s-valid}
iff there is slave's strategy such that for every instance (resp. safe instance) $\Phi$ of $\Psi$, the 
 $\Phi$-play with this 
strategy is successful with respect to every world.  }
\end{definition}

\section{The MALL and MLL fragments}\label{1s6}

Our logic, --- the set of universally valid resources or resource schemata, --- is certainly 
undecidable in the full language as 
it contains first order classical logic. However, some reasonably efficient heuristic
algorithms can apparently be found for it. Also, some natural fragments of the logic are 
decidable. This paper doesn't address these issues in detail as its main goal
is to introduce the language and the semantics and show possible applications in case  
efficient algorithms are elaborated. This is a beginning of the work rather than a completed 
work.

Here we only state the decidability of two fragments of the logic. The first
one we call the MALL fragment. Its  language is the same as that of
Multiplicative-Additive Linear Logic, where $\rii$, $\rcc$, $\acc$ and $\auu$ 
correspond to the multiplicative implication, multiplicative conjunction, additive conjunction
and (additive) universal quantifier of linear logic, respectively.
Here is the definition:

\ 

{\em MALL-formulas}\index{MALL} are the elements of the smallest class $M$ of expressions
such that, saying ``$\Phi$ is a MALL-formula" for $\Psi\in M$, we have:
\begin{itemize}
\item $\rff$ is a MALL-formula;
\item if $\Psi$ is an $n$-ary resource letter and $t_1,...,t_n$ are terms,
then $\Psi(t_1,...,t_n)$ is a MALL-formula;
\item if $\Psi$ and $\Phi$ are MALL-formulas, then so is $\Phi\rii\Psi$;
\item if $\Psi$ and $\Phi$ are MALL-formulas, then so is $\Phi\rcc\Psi$;
\item if $\Psi$ and $\Phi$ are MALL-formulas, then so is $\Phi\acc\Psi$;
\item if $\Psi$ is a MALL-formula and $x$ is a variable, then $\auu x\Psi$ is a 
MALL-formula.\end{itemize}
 
Here is our main technical claim:

\begin{claim} \label{1cl1}
The set of universally s-valid closed MALL formulas is decidable. In particular,
it is the logic $ET$ introduced in \cite{Jap97}.
The decision algorithm is 
constructive: it not only states the existence of a successful strategy (when it exists),
but actually finds such a strategy.\end{claim}

We let this claim go without a proof. An interested reader who carefully
studies the relevant parts of \cite{Jap97}
should be able to re-write the soundness and completeness proof for $ET$
given there as a proof
of the above claim. In fact, the proof given there
establishes the completeness of $ET$ in a much stronger sense than
claimed above.

\ 

A {\em MLL-formula}\index{MLL} (``Multiplicative Linear Logic") is a MALL-formula which does not
contain $\acc$ or $\auu$. Since we have no quantifiers, we assume that 
 all resource letters in  MLL-formulas are 
0-ary. We have a stronger soundness/decidability result for the MLL-fragment
of our logic. Stronger in the sense that it is about validity rather than s-validity. 

A MLL-formula is said to be a {\em binary tautology}, if it is an instance of
a classical tautology (with the double-dots placed over $\fff$, $\fcc$ and $\fii$) in which
every predicate letter (non-$\bot$ propositional atom) occurs at most twice. For example, $\Phi\rcc\Psi\rii\Phi$ is a 
binary tautology, and so is $\Phi\rcc\Phi\rii\Phi$ as the latter is an instance of 
the former; however, $\Phi\rii\Phi\rcc\Phi$ is not a binary tautology. Note that in fact a binary 
tautology is always an instance of a classical tautology where every predicate letter
has either one occurrence, or two occurrences,  one of which is positive and the 
other --- negative. \index{Blass,A.}Blass \cite{Blass} was the 
first to study binary tautologies and find a natural semantics for them.

\begin{claim} A MLL-formula is universally valid iff it is a binary tautology.
Hence, validity for MLL-formulas is decidable; again, the decision algorithm
is constructive.\end{claim}

The ``only if" part of this claim follows from Claim \ref{1cl1} together with an observation
that a MLL-formula is a binary tautology iff it is in $ET$. The ``if" part, as always,
is easier to verify, and instead of giving an actual proof, we will just explain the idea
behind it on particular examples. 

The simplest binary tautology is $\Phi\rii\Phi$. Why is it universally valid? Observe that
since one of the two occurrences of $\Phi$ is negative and the other occurrence is 
positive,  
what is a master's move in one occurrence of $\Phi$, is a slave's move in 
the other occurrence of $\Phi$, and vice versa. The slave's strategy which ensures 
that every play is successful, consists in {\em pairing}\label{1pairing} these two occurrences: copying
master's moves, made in either occurrence, into the other occurrence. For example,
let $\Phi$ be 
\[\alpha\gg\Bigl(\ll(\beta\gg,\ \gamma\gg),\ \ \ll(\delta\gg)\Bigr).\]
Then, the initial position is 
\[\alpha\gg\Bigl(\ll(\beta\gg,\ \gamma\gg),\ \ \ll(\delta\gg)\Bigr)\ \ \rii\ \ 
\alpha\gg\Bigl(\ll(\beta\gg,\ \gamma\gg),\ \ \ll(\delta\gg)\Bigr).\]
Slave waits (keeps returning the above position without changes) until master
makes a move. If master never makes a move, then (after compactization) we deal 
with a one-position (0-move) play and, according to the clauses 2 and 1 of Definition \ref{1d8}, the process produced 
by this play is $\alpha\pii\alpha$. Clearly, this process is valid and hence
the play is successful. Otherwise, if master makes a move at some moment $t_1$, this 
should be replacing the positive occurrence of 
$\alpha\gg\Bigl(\ll(\beta\gg,\gamma\gg),\ \ll(\delta\gg)\Bigr)$ by either $\ll(\beta\gg,
\gamma\gg)$ or
$\ll(\delta\gg)$. Suppose the former. Thus, the next position of the play is 
\[\alpha\gg\bigl(\ll(\beta\gg,\ \gamma\gg),\ \ \ll(\delta\gg)\bigr)\ \ \rii\ \
\ll(\beta\gg,\ \gamma\gg).\]
Then slave does the same in the antecedent of this position, thus making
\[\ll(\beta\gg,\ \gamma\gg)\ \ \rii\ \ 
\ll(\beta\gg,\ \gamma\gg)\]
the next position of the play. This move will happen at time moment $t_2$ which is
greater than $t_1$ by  the time slave needs to copy the master's move.

After this, slave waits again until master reports an execution of this command. If 
this never happens, then the play is successful because, by the clauses 2 and 5 of Definition \ref{1d8}, 
the process it produces is $\pff\pii\pff$, which is always true. Otherwise, at some 
moment $t_3>t_2$, master reports an execution by replacing $\ll(\beta\gg,\ \gamma\gg)$
by, say, $\gamma\gg$. So that the next position now becomes
\[\gamma\gg\ \rii\ \ll(\beta\gg,\gamma\gg).\]
Then, as soon as he can, --- at some moment $t_4$, --- slave reports the same in the consequent 
of this position, and we get
\[\gamma\gg \ \rii\ \gamma\gg.\]
Since there are no moves for this position, the play ends here. An analysis of the clause 6 of Definition
\ref{1d8} convinces us that the process produced by this play is 
\[\Bigl(\alpha\leadsto_{t_2}^{t_3}\gamma\Bigr)\ \pii\ \Bigl(\alpha\leadsto_{t_1}^{t_4}\gamma
\Bigr).\]
Since $t_1<t_2<t_3<t_4$, it is easily seen that this process is valid, and thus 
the play is successful.

A similar strategy can be used for other binary tautologies. E.g., the strategy for 
$\Phi\rcc(\Phi\rii\Psi)\ \rii\ \Psi$ consists in pairing the two occurrences of $\Phi$
 and pairing the two occurrences of $\Psi$. This trick, however, fails for resources
that are not binary tautologies. For example, in $\Phi\rii\Phi\rcc\Phi$, slave can 
pair the negative occurrence of $\Phi$ only with one of the two positive occurrences 
of $\Phi$. The (sub)play over the unpaired occurrence then may produce a false 
process while the (sub)play over the 
negative occurrence --- a true process. In that case the process produced by the whole 
play over  $\Phi\rii\Phi\rcc\Phi$ will be false.

\section{The assembly world in terms processes}\label{1s7}

What is going on in a computer at the assembly language (as well as higher) level 
can be formalized in our language as a process. Here we consider an example of  
a simplified assembly world.

Our ``computer" has only 3 memory locations or registers: $L_1$, initially containing 2, $L_2$, initially 
containing 0, and  $L_3$, initially
containing 0.  The assembly language for this ``architecture" has only 3 commands: command \#1,
 command \#2, command \#3. Command \#i results in 
adding the contents of the other two locations and writing it in $L_i$. 

There are several ways to formalize this situation in our language, and here is one of them.

Since giving a command is an instantaneous event, we assume that command \#i creates 
a ``puls" which makes the contents of $L_i$ change. Creating a puls means making a
certain fact --- denote it by $P_i$ for command \#i --- become true for one  
moment. Between commands another fact, $Np$ (``no puls"), holds.
It can be defined by 
 \begin{equation} \label{1e1}
Np=_{df} \neg  \bigl(P_1\fdd P_2\fdd P_3\bigr).\end{equation}

The further abbreviations we will use for convenience
are:
\begin{eqnarray*}\label{1e2}
\alpha_1=_{df}\ \angle P_1 \pcc\updownarrow Np\\
\alpha_2=_{df}\ \angle P_2 \pcc \updownarrow Np\\
\alpha_3=_{df}\ \angle P_3 \pcc \updownarrow Np   \end{eqnarray*}

 Thus, issuing command \#i can be seen as starting (switching to) process
$\alpha_i$. 

The formulas $\lambda_1$, $\lambda_2$ and $\lambda_3$, defined below, are meant to
describe the behavior
 of the processes going on in the 3 locations:\footnote{
Although our language does not allow terms such as $x+y$, we can pretend that
it does, because every expression containing this kind of terms can be rewritten as an 
equivalent legal expression the language defined in the previous sections. So that, for 
convenience,  here and later we assume 
that our language is based on predicate logic with function symbols and identity 
rather than pure predicate logic.}

\begin{eqnarray*}\label{1e3}
\lambda_1=_{df}\ [\unrhd]\Bigl(\angle P_1\ \pcc\ \updownarrow\fnn P_1\ \pcc\ \pee x,y\bigl(\angle
 L2(x)\pcc\angle L3(y)\pcc\Uparrow 
L1(x+y)\bigr)\Bigr)\\
\lambda_2=_{df}\ [\unrhd]\Bigl(\angle P_2\ \pcc\ \updownarrow\fnn P_2\ \pcc\ \pee x,y\bigl(\angle
 L1(x)\pcc\angle L3(y)\pcc\Uparrow 
L2(x+y)\bigr)\Bigr)\\
\lambda_3=_{df}\ [\unrhd]\Bigl(\angle P_3\ \pcc\ \updownarrow\fnn P_3\ \pcc\ \pee x,y\bigl(\angle
 L1(x)\pcc\angle L2(y)\pcc\Uparrow 
L3(x+y)\bigr)\Bigr)\end{eqnarray*}

Before we analyze the $\lambda_i$, let us agree on some jargon. Every (true) process $[\rhd]\gamma$
or $\gamma_1\rhd...\rhd\gamma_n$ can be divided into $\rhd$-{\em stages}, which are the consecutive
time intervals on which $\gamma$ is true. A transition from one stage 
to another will be referred to as a {\em $\rhd$-transition}. Similarly, we will use 
the terms ``$\unrhd$-stage" and ``$\unrhd$-transition" for processes of the type $[\unrhd]\gamma$
or $\gamma_1\unrhd...\unrhd\gamma_n$.

In these terms, $\lambda_i$ starts its $\unrhd$-stage when puls $P_i$ is given (the conjunct $\angle
P_i$), and will stay in this stage exactly until the same puls is given again. 
 A $\unrhd$-transition to the new stage before this is impossible 
because, due to the conjunct $\angle P_i$, that stage requires that $P_i$ be true 
at the moment of the transition. And a late transition to a new
stage is also impossible because, as soon as $P_i$ becomes true, the conjunct $\Uparrow\fnn
P_i$ is violated.  Throughout each $\unrhd$-stage, except its first moment, the location
$L_i$ then stores the sum of the values that the other two locations 
had at the initial moment of the stage.

Now we need to axiomatize the situation
where the initial value of $L1$ is 2,  the initial values of the other two locations 
are 0, and these values will be maintained until the corresponding command (puls) is
given. This will be captured by the following 3 axioms:

\begin{eqnarray}
\label{1e4}
\Bigl(\bigl(\Updownarrow L_1(2)\bigr)\pcc\bigl(\Downarrow\neg P_1\bigr)\Bigr)\unrhd\lambda_1
\end{eqnarray}

\begin{eqnarray}\label{1e5}
\Bigl(\bigl(\Updownarrow L_2(0)\bigr)\pcc\bigl(\Downarrow\neg P_2\bigr)\Bigr)\unrhd\lambda_2
\end{eqnarray}

\begin{eqnarray}\label{1e6}
\Bigl(\bigl(\Updownarrow L_3(0)\bigr)\pcc\bigl(\Downarrow\neg P_3\bigr)\Bigr)\unrhd\lambda_3
\end{eqnarray}

Next, for safety, we need to state that  two different pulses cannot happen simultaneously:
\begin{eqnarray}\label{1e7}
\Box \Bigl(\neg(P_1\fcc P_2)\fcc\neg(P_1\fcc P_3)\fcc\neg(P_2\fcc P_3)\Bigr)
\end{eqnarray}

We also need to axiomatize some sufficient amount of the arithmetic needed. We may assume
that $Arithm$ is the conjunction of the axioms of Robinson's arithmetic (see \cite{Kleene}), 
although, for our purposes, just 
\[2+0=2\ \fcc \ 2+2=4\ \fcc\ 2+4=6\ \fcc\ 6+4=10\]  would do as 
$Arithm$. In any case, 
\begin{equation}\label{1e7.1}
\Box Arithm\end{equation}
should be one of our axioms.

The final axiom below represents a program which, after being idle ($\Downarrow Np$),
issues command \#2, then command \#3, then command \#1 and then, again, command  \#2.

\begin{equation}\label{1e8}
(\Downarrow Np)\rhd\alpha_2\rhd\alpha_3\rhd\alpha_1\rhd\alpha_2. 
\end{equation}

Our claim is that given the truth of these axioms, we can conclude that the process 
$\Updownarrow L_2(10)$ will be reached at some point. In other words, the process

\[\Bigl((\ref{1e4})\pcc(\ref{1e5})\pcc(\ref{1e6})\pcc(\ref{1e7})\pcc(\ref{1e7.1})\pcc(\ref{1e8})
\Bigr)\pii\Bigl(\ptt\rhd\Updownarrow L_2(10)\Bigr)\]
is valid. Indeed, in the initial situation,\footnote{
We use the word ``situation" with a relaxed meaning: here it denotes some ``core" subset 
of facts rather than a complete set of facts.}
 we have \[L_1(2),\ L_2(0),\ L_3(0),\ Np.\] 
While we have $Np$, the values of $L_1$, $L_2$, $L_3$ cannot change, because a $\unrhd$-transition
to $\lambda_i$ would require the truth (at the moment of transition) of $P_i$, which is ruled out 
by (\ref{1e1}). So, the situation will change only if a puls $P_i$ occurs.

The first stage $\Downarrow Np$ of axiom \ref{1e8} prevents occurring such pulses. So, the 
situation will change exactly when the process (\ref{1e8}) makes a $\rhd$-transition
 from $\Downarrow Np$ to 
$\alpha_2$, i.e. to $\angle P_2\pcc\updownarrow Np$. This transition forces (\ref{1e5}) to switch 
to $\lambda_2$, which results in starting the process $\Uparrow L_2(2)$: $L_2$ will have
its old value $0$ at the first moment of the stage, and  the value $2$ after that. On the other hand, in 
view of (\ref{1e7}), no $\unrhd$-transition can happen
in (\ref{1e4}) or (\ref{1e6}). Thus, the situation at the moment of transition becomes 
\[ L_1(2),\ L_2(0),\ L3(0),\ P_2,\]
which will become 
\[ L_1(2),\ L_2(2), \ L3(0),\  Np\]
right after the moment of transition because of $\Uparrow L_2(2)$ and $\updownarrow Np$. 

Continuing arguing in this manner, we get that the further development of situations
is:
\[ L_1(2), \ L_2(2), \ L_3(0),\ P_3,\]
\[ L_1(2), \ L_2(2), \ L_3(4),\ Np,\]
\[L_1(2), \ L_2(2), \ L_3(4), \ P_1,\]
\[L_1(6), \ L_2(2), \ L_3(4), \ Np,\]
\[L_1(6), \ L_2(2), \ L_3(4), \ P_2,\]
\[L_1(6), \ L_2(10), \ L_3(4), \ Np.\]

Since the last stage of the program (\ref{1e8}) contains the conjunct $\updownarrow Np$, no further 
changes will occur, and the value of $L_2$ will remain 10.

\section{The assembly world in terms of resources}\label{1s8}

In the previous example we dealt with processes that ran ``by themselves". We  could
not interfere and manage them, so that there was no space for planning.

Presenting the world as a set of resources  rather than processes allows us 
to capture our ability to influence the course of events in the world. Our way 
to interact with the world is giving and receiving commands.

Here is an attempt to present the assembly world as a resource. We assume that 
we have, as an empty-potential resource,  
the $\pcc$-conjunction $\Gamma$ of the axioms (\ref{1e4})-(\ref{1e7.1}), suffixed by $\gg$:
\[\Gamma=_{df}\ \Bigl((\ref{1e4})\pcc(\ref{1e5})\pcc(\ref{1e6})\pcc(\ref{1e7})\pcc(\ref{1e7.1})
\Bigr)\gg.\]
As for axiom (\ref{1e8}), 
which is a ``ready program", instead of it we have
a resource which accepts from us any of those 3 commands, as many times as we like.

This resource will be expressed by \begin{equation}\label{1e9}
\bigl(\Downarrow Np\bigr)\Theta,\end{equation}
 where $\Theta$  is introduced by the recursive definition
\[\Theta:=\ \gg\Bigl(\ll(\alpha_1\Theta),\ll(\alpha_2\Theta),\ll(\alpha_3\Theta)\Bigr). 
\]

Now we can ask the question if we can accomplish the task $ 
\Bigl(\ptt\rhd\bigl(\Updownarrow L2(10)\bigr)\Bigr)\gg$. In other words, whether 
\[\Gamma\rcc(\ref{1e9})\ \rii\ \Bigl(\Bigl(\ptt\rhd\bigl(\Updownarrow L2(10)\bigr)\Bigr)\gg\
\Bigr)\]
is universally valid.
Yes, it is. The strategy is:

\ 

 Convert $(\Downarrow Np)\Theta$ into 
$\ll(\alpha_2\Theta)$; after getting the report
 $\alpha_2\Theta$,
convert it into 
$\ll(\alpha_3\Theta)$; after getting the report
 $\alpha_3\Theta$,
convert it into 
$\ll(\alpha_1\Theta)$; after getting the report
 $\alpha_1\Theta$,
convert it into 
$\ll(\alpha_2\Theta)$, and stop.

\ 

Thus, unless (\ref{1e9}) fails to carry out its task, the only play
corresponding 
to this strategy is the following sequence of positions:

$0$. $\Gamma\rcc \bigl( (\Downarrow Np)\Theta\bigr)\ \rii\ 
\Bigl(\Bigl(\ptt\rhd\bigl(\Updownarrow L2(10)\bigr)\Bigr)\gg\Bigr)$

$1$. $\Gamma\rcc \bigl(\ll \alpha_2\Theta\bigr)\ \rii\ 
\Bigl(\Bigl(\ptt\rhd\bigl(\Updownarrow L2(10)\bigr)\Bigr)\gg\Bigr)$

$1'$. $\Gamma\rcc \bigl(\alpha_2\Theta\bigr)
\ \rii\ \Bigl(\Bigl(\ptt\rhd\bigl(\Updownarrow L2(10)\bigr)\Bigr)\gg\Bigr)$

$2$. $\Gamma\rcc \bigl(\ll\alpha_3\Theta\bigr)
\ \rii\ \Bigl(\Bigl(\ptt\rhd\bigl(\Updownarrow L2(10)\bigr)\Bigr)\gg\Bigr)$

$2'$. $\Gamma\rcc \bigl(\alpha_3\Theta\bigr)
\ \rii\ \Bigl(\Bigl(\ptt\rhd\bigl(\Updownarrow L2(10)\bigr)\Bigr)\gg\Bigr)$

$3$. $\Gamma\rcc \bigl(\ll \alpha_1\Theta\bigr)
\ \rii\ \Bigl(\Bigl(\ptt\rhd\bigl(\Updownarrow L2(10)\bigr)\Bigr)\gg\Bigr)$

$3'$. $\Gamma\rcc \bigl( \alpha_1\Theta\bigr)
\ \rii\ \Bigl(\Bigl(\ptt\rhd\bigl(\Updownarrow L2(10)\bigr)\Bigr)\gg\Bigr)$

$4$. $\Gamma\rcc \bigl(\ll \alpha_2\Theta\bigr)
\ \rii\ \Bigl(\Bigl(\ptt\rhd\bigl(\Updownarrow L2(10)\bigr)\Bigr)\gg\Bigr)$

$4'$. $\Gamma\rcc\bigl(\alpha_2\Theta\bigr)
\ \rii\ \Bigl(\Bigl(\ptt\rhd\bigl(\Updownarrow L2(10)\bigr)\Bigr)\gg\Bigr)$.

\ 

One can easily see that this play produces the same process as the 
process described in the previous section, and hence it achieves 
the goal provided that the resources $\Gamma$ and (\ref{1e9}) successfully accomplish
their tasks. 

\ 

However, this is not the best way to represent the assembly world. 
Although it avoids the representational frame problem, --- no need in anything
like frame axioms, --- the inferential
frame problem\index{frame problem}\footnote{For a discussion of these 2 sorts of frame problem, see 
\cite{Russel}.} still remains: in an analysis of the play, after every
move, we need to verify that only one $\pcc$-conjunct of $\Gamma$ 
(which is the major part of the world) changes its $\unrhd$-stage; these changes are not 
reflected in the current position --- $\Gamma$ remains $\Gamma$  
and we need to separately keep track of in what stages the $\unrhd$-processes of its effect are.
It is just the presence of the operators $\rhd$, $\unrhd$, $[\rhd]$, $[\unrhd]$ in 
resources that makes a trouble of this kind. 

A better way to represent the assembly world, which avoids the inferential (along 
with the representational) frame problem  is to view each memory location as an independent
resource which can directly accept our commands. Then we can be sure that a command 
changes only the 
resource (the contents of the location) to which it is given, and we don't need to check 
or re-write the other agents of the system as long as their effects don't contain
the operators $\rhd$, $\unrhd$, $[\rhd]$, $[\unrhd]$.

Below  is a description of this sort of axiomatization for the assembly world. Observe 
that it is totally $(\rhd,\unrhd,[\rhd],[\unrhd])$-free.

 Let

\[\Lambda_1:=\ \gg\ll\Bigl(\Bigl(\pee x,y\bigl(\angle L_2(x)\pcc\angle L_3(y)\pcc\Uparrow 
L_1(x+y)\bigr)\Bigr)\Lambda_1\Bigr)
\]
\[\Lambda_2:=\ \gg\ll\Bigl(\Bigl(\pee x,y\bigl(\angle L_1(x)\pcc\angle L_3(y)\pcc\Uparrow 
L_2(x+y)\bigr)\Bigr)\Lambda_2\Bigr)
\]
\[\Lambda_3:=\ \gg\ll\Bigl(\Bigl(\pee x,y\bigl(\angle L_1(x)\pcc\angle L_2(y)\pcc\Uparrow 
L_3(x+y)\bigr)\Bigr)\Lambda_3\Bigr)
\]
We assume the following axioms:
\begin{equation}\label{1e14}
\Bigl(\Updownarrow L_1(2)\Bigr)\Lambda_1
\end{equation}
\begin{equation}\label{1e15}
\Bigl(\Updownarrow L_2(0)\Bigr)\Lambda_2
\end{equation}
\begin{equation}\label{1e16}
\Bigl(\Updownarrow L_3(0)\Bigr)\Lambda_3
\end{equation}
together with $\Bigl((\ref{1e7})\rcc(\ref{1e7.1})\Bigr)\gg$. 

Thus, each of the agents (\ref{1e14}), (\ref{1e15}), (\ref{1e16}) accepts one single
command, the execution of which results in writing in the corresponding 
location the sum of 
the contents of the other two locations. A strategy for achieving the goal 
$\bigl(\ptt\rhd\Updownarrow L2(10)\bigr)\gg$ is: Give a command to (\ref{1e15}), then to 
(\ref{1e16}), then to (\ref{1e14}) and then, again, to (\ref{1e15}). A reasonable algorithm 
which finds this kind of strategy and verifies its successfulness, would only keep 
track of the changes that occur in the effect of the resource to which a command 
is given. As we noted, however, this relaxed behavior of the algorithm is possible 
only if those effects don't contain the 
``trouble maker" operators $\rhd$, $\unrhd$, $[\rhd]$ and $[\unrhd]$.

\end{document}